\documentstyle[eqsecnum,aps,prb,two column,epsf]{revtex} %

\makeatletter

\def\inc@eqnnum{\addtocounter{equation}{1}}
\def\dec@eqnnum{\addtocounter{equation}{-1}}
\@definecounter{equation}
\ifsecnumbers
 \@addtoreset{equation}{section}
 \def\theequation@prefix{\arabic{section}.}
\else
 \def\theequation@prefix{}
\fi

\makeatother

\begin{document} 

\title{Magnetic, thermal, and transport properties on single crystals of
antiferromagnetic Kondo-lattice Ce$_{2}$PdSi$_{3}$}
 
\author{S. R. Saha, H. Sugawara, T. D. Matsuda, Y. Aoki, and H. Sato}
 
\address{Department of Physics, Tokyo Metropolitan University,
Hachioji-Shi, Tokyo 192-0397, Japan}

\author{E. V. Sampathkumaran}
\address{Tata Institute of Fundamental Research, Homi
Bhabha Road, Colaba, 400005 Mumbai, India}
\date{\today}

\maketitle
 
\begin{abstract}
Magnetization, heat capacity, electrical resistivity,
thermoelectric power, and Hall effect have been investigated on
single-crystalline Ce$_{2}$PdSi$_{3}$. This compound is shown to order 
antiferromagnetically below N\'eel temperature $(T_{N})\sim$ 3 K.
The Sommerfeld coefficient far below $T_{N}$ is found to be about
110 mJ/K$^{2}$ mol Ce, which indicates the heavy-fermion character of this 
compound. The transport and magnetic properties exhibit large anisotropy with
an interplay between crystalline-electric-field (CEF) and Kondo effects. The
sign of thermoelectric power is opposite for different directions at high 
temperatures and the ordinary Hall coefficient is anisotropic with opposite
sign for different geometries, indicating the anisotropic Fermi surface.
The CEF analysis from the temperature dependence of magnetic
susceptibility suggests that the ground state is $\mid\pm\frac{1}{2}\rangle$.
The first and the second excited CEF doublet levels are found to be located at
about 30 and 130 K, respectively. The Kondo temperature is estimated to be
the same order as $T_N$, indicating the presence of a delicate competition
between the Kondo effect and magnetic order.
\end{abstract}
\vskip 0.5cm
PACS number(s): 75.30.Mb, 71.27.$+$a, 71.70.ch, 75.40.$-$s

\section{Introduction}
There has been considerable interest in understanding the interplay among
the crystalline-electric-field (CEF) effect, the indirect exchange
[Ruderman-Kittel-Kasuya-Yoshida (RKKY)]
interaction among the 4$f$ magnetic moments and the Kondo effect in Ce
compounds, since these are the decisive factors of the physical properties
in these compounds. It is therefore worthwhile to carry out careful
investigation in new Ce compounds. With this motivation, we report here the
results of magnetic susceptibility ($\chi$), magnetization ($M$),
heat-capacity ($C$), electrical-resistivity ($\rho$), thermoelectric-power
($S$), and Hall-coefficient ($R_{H}$) measurements on single-crystalline
Ce$_{2}$PdSi$_{3}$, grown for the first time.

This compound has been reported to form in an AlB$_{2}$-derived hexagonal
crystal structure and to exhibit Kondo effect.\cite{1} The intensity of
investigation in the RE$_{2}X$Si$_{3}$ (RE $=$ rare earth, $X$ $=$ transition
metal) series, crystallizing in the above-mentioned structure,\cite{2}
increased only in the recent years and these compounds have been reported 
to exhibit many unusual features in the magnetic, thermal, and transport
properties (see, for instance, Refs. 1--10 and references therein).
Gd$_{2}$PdSi$_{3}$ exhibits Kondo-lattice-like anomalies, e.g., a resistivity
minimum above $T_{N}$ accompanied by a large negative 
magnetoresistance.\cite{3,4} These features, presumably due to a novel
mechanism, are not common to Gd compounds. Ce$_{2}$CoSi$_{3}$ is a
mixed-valent compound, a small La substitution for Ce induces a
non-Fermi-liquid behavior in $\rho$.\cite{5} Eu$_{2}$PdSi$_{3}$ exhibits
two distinct magnetic transitions, with the possibility of
quasi-one-dimensional magnetism for the high-temperature transition\cite{6}
and unusual magnetic characteristics.\cite{7} Particularly considering that
the Gd-based compound in the Pd series has been found to show many interesting
anisotropic features,\cite{3} it is tempting to carry out detailed
studies on the single-crystalline Ce$_{2}$PdSi$_{3}$ as well. Previous
magnetic, electrical-resistance, and heat-capacity measurements on this
compound were performed only in the polycrystalline form\cite{1} and no
clear magnetic ordering could be detected. Thus the present
studies extended to much lower temperatures, particularly on single crystals, 
serve as a first thorough characterization of the bulk properties of this 
compound.

\section{Experimental details}
Single crystals of Ce$_{2}$PdSi$_{3}$ have
been prepared by the Czochralski pulling method using a tetra-arc furnace
in an argon atmosphere. The single-crystalline nature has been confirmed
by back-reflection Laue technique. The magnetic measurements were carried
out with a Quantum Design superconducting quantum interference device
(SQUID) magnetometer. The heat capacity was measured by a quasiadiabatic heat-
pulse method using a dilution refrigerator. The electrical-resistivity
and Hall-effect measurements have been performed by a conventional dc 
four-probe method in the temperature
interval of 0.5--300 K. The thermoelectric-power data have been taken by the
differential method using a Au-Fe (0.07\%)-chromel thermocouple.

\section{Results and discussions}
Figure 1(a) shows the temperature
dependence of the inverse magnetic susceptibility $\chi$$^{-1}(T)$,
measured in a magnetic field $H=$ 1 kOe for both $H//[10\overline10]$ and
$H//[0001]$. There is a large difference in the absolute values of $\chi$
for two geometries, apparently due to CEF effect. The effective
magnetic moment ($\mu_{eff}$)
and the paramagnetic Curie temperature $\Theta_{P}$ are estimated from the
high-temperature linear region to be about 2.60$\mu_{B}$/Ce
\begin{center}
\begin{figure}
\epsfxsize=8.6cm \epsfbox{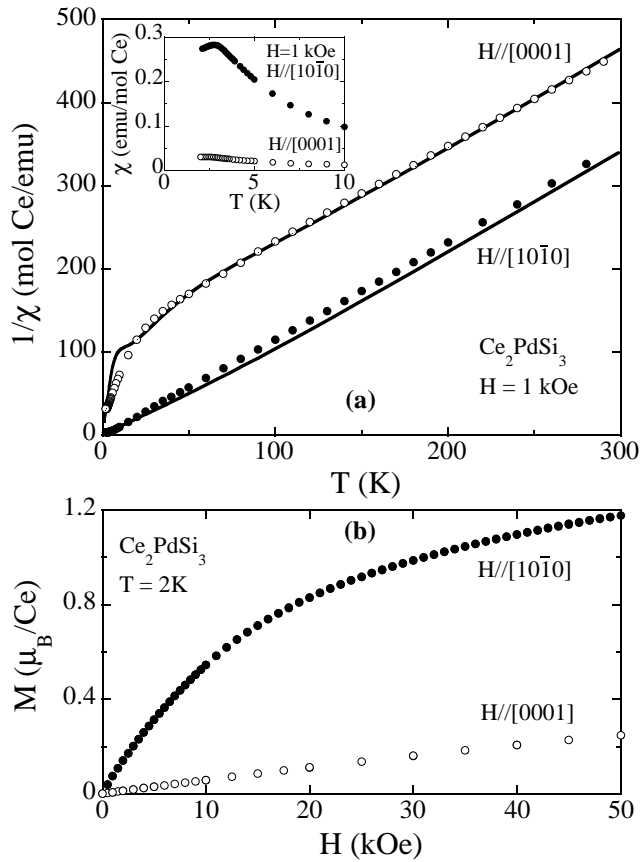}
\caption{(a) The inverse magnetic susceptibility versus temperature (2--300
K) for single crystals of Ce$_{2}$PdSi$_{3}$. The values calculated
considering the CEF model (see text) are shown by the solid
curves. The inset shows the expanded view of the magnetic susceptibility at
low temperatures. (b) The isothermal magnetization behavior at 2 K for
$H//[10\overline10]$ and $H//[0001]$.}
\end{figure} 
\end{center}
and 3.6 K for $H//[10\overline10]$ and 2.65$\mu_{B}$/Ce and
$-$107 K for $H//[0001]$, respectively. These values of $\mu_{eff}$ are very
close to that expected for a free trivalent Ce ion (2.54$\mu_{B}$). The large
negative value of $\Theta_{P}$ for $H//[0001]$ and a small positive value for
$H//[10\overline10]$ are presumably due to CEF effect.\cite{11} The large
anisotropy in $\Theta_{P}$ may also indicate the existence of anisotropy in
the exchange interaction which depends on the CEF level scheme. The expanded
view of the temperature dependence of $\chi$ is shown in the inset of
Fig. 1(a). There is no difference between the field-cooled and the
zero-field-cooled measurements of $\chi$ down to 1.9 K within the experimental
accuracy, indicating the absence of any spin-glass-like behavior. This fact
is in contrast to the formation of spin-glass state in the isostructural
U$_2$PdSi$_3$.\cite{9} There is a peak in $\chi(T)$ at around 2.8 K for
$H//[10\overline10]$ and at 2.5 K for $H//[0001]$. These peaks can be ascribed
to the antiferromagnetic ordering. The occurrence of the peak at slightly
different temperatures for two directions might be due to the anisotropic
field dependence of $T_N$ for two directions, since
the heat-capacity measurement in absence of a magnetic field shows a peak at
around 3 K as shown below. Recent neutron-diffraction data on polycrystals
also suggests the occurrence of antiferromagnetic (AF) ordering below
2.5 K in a sinusoidally modulated AF structure.\cite{10}

We have tried to analyse the $\chi(T)$ data 
using a CEF model, considering hexagonal site symmetry\cite{10} of Ce
in Ce$_{2}$PdSi$_{3}$. According to the Hutchings'
notation,\cite{12} the CEF Hamiltonian for $J=5/2$ ion with the hexagonal
point symmetry is given by
\begin{equation}
{\mathcal{H}}=B_{2}^{0}O_{2}^{0}+B_{4}^{0}O_{4}^{0},
\end{equation}
where $B_{n}^{m}$ and $O_{n}^{m}$
represent the CEF parameters and the Steven's equivalent operators,
respectively. The results of this CEF analysis using Eq. (1),
employing the $\chi(T)$ data at the paramagnetic region, leads
to $B_{2}^{0}\simeq$ 10.1 K, and $B_{4}^{0}\simeq$ 0.11 K. This set of
parameters corresponds to a crystal-field level scheme with the
three doublets $\mid\pm\frac{1}{2}\rangle$, $\mid\pm\frac{3}{2}\rangle$ and
$\mid\pm\frac{5}{2}\rangle$ at around 0, 28 K ($\Delta_{1}$), and 130 K
($\Delta_{2}$), respectively. Accordingly, the calculated values of
$\chi$$^{-1}$ are shown by the solid lines in Fig. 1(a), which indicates that
the anisotropy in $\chi(T)$ is mainly induced by the CEF effect. However,
there is a deviation of the calculated $\chi$$^{-1}$ from the experimental
values, particularly at low temperatures. The following explanations can be
offered to this deviation: According to Szytula {\it et al},\cite{10}
Pd and Si atoms are random in this crystal structure (space group $P6/mmm$).
Therefore this randomness or disorder between Pd and Si sites
may produce local modification of the CEF effects due to the distribution of
the CEF parameters. Alternatively, if Pd and Si are well ordered (space group
$P6_{3}/{mmc}$, see Ref. 10), there are two different
crystallographic environments for Ce ions,\cite{8,13} in which case the CEF
effect may be different for these two sites. All these factors  are neglected
in the present CEF calculations.

The isothermal magnetization at 2 K is shown in Fig. 1(b). $M$ varies
distinctly in different ways with the applied magnetic field for
$H//[10\overline10]$ and $H//[0001]$. This anisotropy in $M$ is presumably
due to the CEF effect. The magnetic moments at $H=$ 50 kOe for two
directions are $\sim$ 1.18 and $\sim$ 0.25$\mu_{B}$/Ce for
$H//[10\overline10]$ and $H//[0001]$, respectively. The larger magnetic
moment for $H//[10\overline10]$ indicates the $a$-$b$ plane as the easy
plane of magnetization, and the $\mid\pm\frac{1}{2}\rangle$ doublet as the
ground state, in agreement with the CEF analysis from $\chi(T)$ described
above. These two facts are also consistent with the proposed magnetic
structure of Ce$_{2}$PdSi$_{3}$ based on neutron-diffraction experiment, i.e.,
the Ce magnetic moments lie in the $a$-$b$ plane.\cite{10}
\begin{center}
\begin{figure} 
\epsfxsize=8.6cm \epsfbox{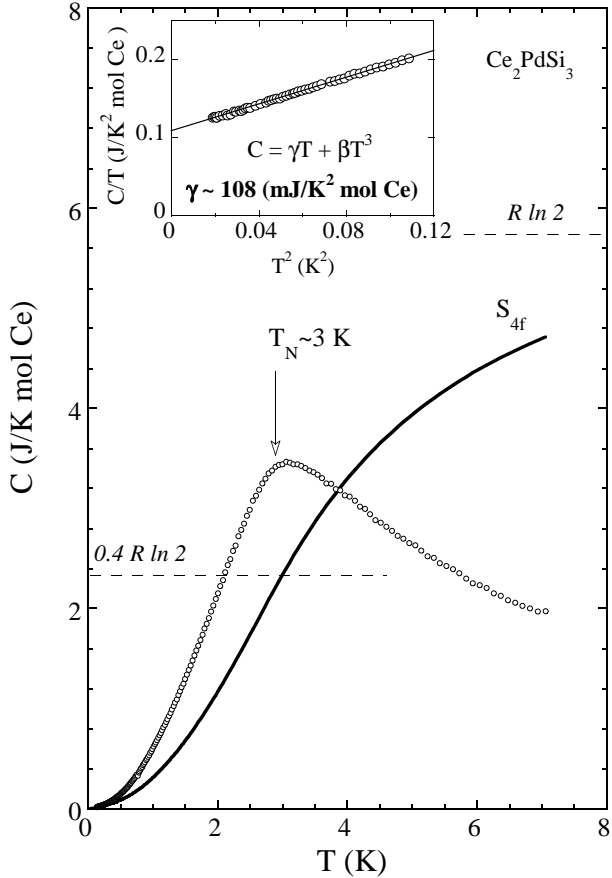}
\caption{The temperature dependence of heat capacity $C(T)$ for
Ce$_{2}$PdSi$_{3}$. The magnetic (4$f$ contribution) entropy is presented by
the solid line. The inset shows the $C/T$ vs $T^2$ plot at
temperatures well below $T_N$ where phonon contribution can be neglected.}
\end{figure} 
\end{center}
Figure 2 shows the temperature dependence of heat capacity, $C(T)$, for
Ce$_{2}$PdSi$_{3}$. A peak in $C(T)$ at around 3 K indicates the existence
of the antiferromagnetic ordering below $T_{N}\simeq$ 3 K, supporting the
conclusion from the $\chi$$(T)$ behavior. The transition is, however, not very
sharp, which might be due to the presence of site disorders. The inset shows
the $C/T$ vs $T^2$ plot at temperatures far below $T_N$ where phonon
contribution is negligible. The linear coefficient of specific heat
($\gamma$), named as the Sommerfeld coefficient, estimated from this plot at
such low temperatures is 108 mJ/K$^{2}$ mol Ce. At this temperature
range the specific heat can be expressed as $C=\gamma T+\beta T^{3}$
with $\beta\simeq$ 862 mJ/K$^{4}$ mol Ce and the parameter $\beta$ mainly comes
from the contribution of an antiferromagnetic-magnon part. For
Y$_{2}$PdSi$_{3}$, the non-magnetic compound serving as a reference for
phonon contribution, the $\gamma$ value is 4.5 mJ/K$^{2}$ molY.\cite{1}
The moderately large value of $\gamma$ for Ce$_{2}$PdSi$_{3}$ suggests that
even in the magnetically-ordered state Ce$_{2}$PdSi$_{3}$ may be classified as 
a heavy fermion. The solid line in Fig. 2 shows the magnetic entropy ($S_{4f}$)
estimated from the 4$f$ contribution ($C_{m}$) to $C$. $C_{m}$ is obtained by
employing the $C$ values of Y$_{2}$PdSi$_{3}$ (Ref. 1) as a reference for the
lattice contribution taking into account the difference of Debye temperatures
of the two compounds by using the procedure suggested in Ref. 14. At $T_{N}$,
$S_{4f}\simeq$ 2.3 J/K mol Ce is $\sim$ 40\% of $R$ ln 2 expected for a
complete removal of the two fold degeneracy of a CEF ground-state doublet.
This reduced entropy value might be due to the substantial Kondo-derived
reduction of the Ce moments and/or a presence of short-range correlations
above $T_{N}$.\cite{15} Tentatively assuming a Kondo-derived reduction, the
magnetic entropy $S_{4f}$ ($\simeq$ 0.4 $R$ ln 2) at $T_{N}$ yields the Kondo
temperature
$T_{K}\simeq$ 8 K, according to the Bethe-ansatz for a spin-$\frac{1}{2}$
Kondo model (see Refs. 16 and 17). The theoretical calculations
of $C(T)$ using $T_{K}\simeq$ 8 K, however, cannot reproduce the experimental
curve above $T_{N}$. This deviation indicates the presence of
short-range-AF correlations above $T_{N}$; the presence of these
correlations is, in fact, detected in a recent neutron-scattering
experiment,\cite{18} and a possible contribution from an excited CEF
level (28 K) even at the measured temperature range.

The temperature dependence of resistivity $\rho(T)$ for Ce$_{2}$PdSi$_{3}$
with the current $J//[10\overline10]$ and $J//[0001]$ as well as in
polycrystalline Y$_{2}$PdSi$_{3}$ is shown in Fig. 3. $\rho$
for both current directions gradually decreases with decreasing
temperature down to about 20 K showing a broad hump around 100 K; below
20 K, there is a weak upturn giving rise to a minimum at around 20 K,
followed by a drop below 8 K. The $\rho(T)$ in Y$_{2}$PdSi$_{3}$ shows usual
metallic behavior, however there is a drop in $\rho$ below 6 K, presumably
attributable to the presence of traces of the superconducting phase
YPdSi.\cite{1} As known for many other Ce alloys, the broad hump in
Ce$_{2}$PdSi$_{3}$ can be ascribed to the combined effect of CEF and Kondo 
effect. The magnetic contribution to the resistivity
$\rho_{m}$$=\rho$(Ce$_{2}$PdSi$_{3}$)$-\rho$(Y$_{2}$PdSi$_{3}$), obtained by
using $\rho(T)$ data for $J//[0001]$ in Ce$_{2}$PdSi$_{3}$, is shown
in the inset; the ratio of the slopes of the $\rho_{m}$ vs ln $T$
plot at high to low temperature turns out close to 35/3 
confirming that the ground state is a doublet (see Ref. 19). It clearly
reveals the presence of a peak at around 50 K which
might be related with $T_{K}$ enhanced by CEF effect as proposed by Hanzawa
{\it et al}.\cite{20} Correspondingly, there is also a broad hump in
$S(T)$ (see Fig. 4). The minimum in $\rho(T)$ around 20 K is either due to
the Kondo effect or a consequence of magnetic-precursor effect as noted for
some Gd alloys.\cite{4} It may be noted that the drop in $\rho$ sets in at 8 K,
much above the $T_{N}$, similar to CePt$_{2}$Ge$_{2}$,\cite{21}
\begin{center}
\begin{figure} 
\epsfxsize=8.6cm \epsfbox{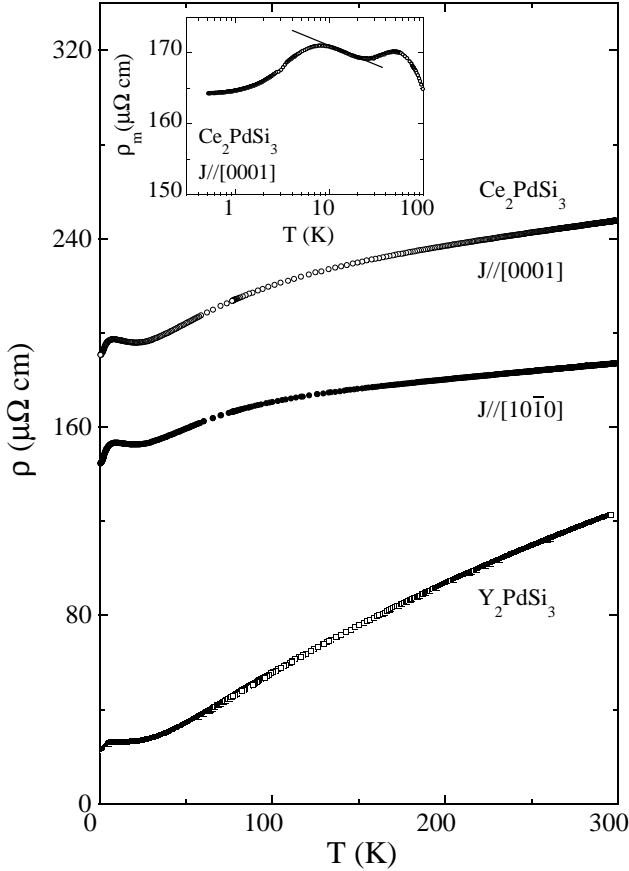}
\caption{The electrical resistivity ($\rho$) of single-crystalline
Ce$_{2}$PdSi$_{3}$ as a function of temperature (0.5--300 K) for
$J//[10\overline10]$ and $J//[0001]$ and polycrystalline Y$_{2}$PdSi$_{3}$.
The inset shows the magnetic contribution to the resistivity as a function of
ln $T$.}
\end{figure}
\end{center}
and such a feature in magnetically ordered Kondo lattices arises from a
combination of indirect exchange interaction and the Kondo effect.\cite{22}
There is a small difference in absolute values for two geometries,
which might be due to the combined effect of anisotropy in the Fermi surface
and preferably oriented microcracks. It may also be remarked that
the residual resistivity is large even for the single crystal, which might be
due to a presence of crystallographic (Pd-Si) disorder or a combined
effect of disorder and a dominance of Kondo contribution even in the
magnetically ordered state.

Figure 4 shows the temperature dependence of thermoelectric power ($S$) in
Ce$_{2}$PdSi$_{3}$ as well as in Y$_{2}$PdSi$_{3}$ (polycrystal). In
Ce$_{2}$PdSi$_{3}$, for the temperature gradient $\Delta$$T//[10\overline10]$,
$S$ is positive at room temperature, then it gradually increases with
decreasing temperature and shows a broad hump around 100 K. There is a change
of sign around 50 K with a minimum around 17 K. On the other hand, for
$\Delta$$T//[0001]$, $S$ has a large negative value at room temperature and
decreases with decreasing
\begin{center}
\begin{figure} 
\epsfxsize=8.6cm \epsfbox{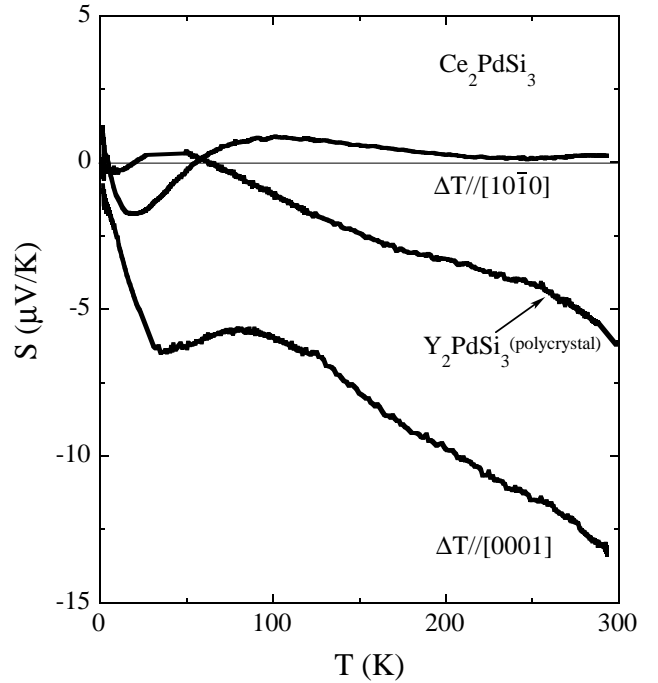}
\caption{The thermoelectric power as a function of temperature for two
different directions of thermal gradient in Ce$_{2}$PdSi$_{3}$ single crystals,
along with the data for polycrystalline Y$_{2}$PdSi$_{3}$.}
\end{figure}
\end{center}
temperature showing a broad hump around 100 K.
$S(T)$ for this direction also shows a minimum, however, at a temperature
slightly higher than that for $\Delta$$T//[10\overline10]$. Thus $S(T)$ in
Ce$_{2}$PdSi$_{3}$ is highly anisotropic with the directions of the thermal
gradient. The anisotropy in the Fermi surface might be one reason behind this
anisotropy, since $S(T)$ in the isostructural Gd$_{2}$PdSi$_{3}$ is also
anisotropic.\cite{3} For nonmagnetic Y$_{2}$PdSi$_{3}$, $S(T)$ has a large
negative value at room temperature and decreases gradually with temperature.
In Ce$_{2}$PdSi$_{3}$, the broad hump can be attributed to the interplay
between CEF and Kondo effect.\cite{23} The origin of the minimum at
$T_{min}$ $\simeq$ 17 K is not clear yet, however, the possible explanation
may be the Kondo scattering in the CEF ground state or the growth of AF
correlations as in the case of CeAuAl$_{3}$.\cite{24} Tentatively assuming the
Kondo-derived origin, $T_{K}$ $\simeq$ 8 K would be obtained using the relation
$T_{K}$ $\simeq$ $\frac{1}{2}T_{min}$ that holds for CeAl$_{2}$ and CeCu$_2$
(see Ref. 15). However, the temperature dependence of $C$ above $T_{N}$
suggests that this estimation of $T_{K}$ is a rough one, indicating that both
Kondo effect and AF correlations may play a role for this minimum.
For $\Delta$$T//[10\overline10]$, $S(T)$ is similar to the behavior
in the typical magnetically ordered heavy-Kondo compounds, e.g., CeCu$_{2}$ and
CeAl$_{2}$,\cite{15} though the overall temperature dependence of $S$ in
Ce$_{2}$PdSi$_{3}$ is rather weaker. If the crystallographic disorder is the
dominant origin of the large residual resistivity, the Kondo contribution to
$S(T)$ could be suppressed. Since, according to the Gorter-Nordheim rule, the
thermoelectric power for more than one scattering mechanisms can be expressed
as $S_{alloy}=[\rho_{1}S_{1}+\rho_{2}S_{2}]/\rho$, where the subscripts 1 and 2
correspond to different scattering mechanism:\cite{25} In the present case,
1 represents the Kondo scattering and 2 represents the other scatterings.
Therefore a large $\rho_{2}$ can suppress the Kondo contribution $S_{1}$. The
large negative thermoelectric power in Y$_2$PdSi$_3$ might arise from the 
4$d$ band of Pd, as in the case of the 3$d$ band of Co in YCo$_{2}$.\cite{26}
\begin{center}
\begin{figure} 
\epsfxsize=8.6cm \epsfbox{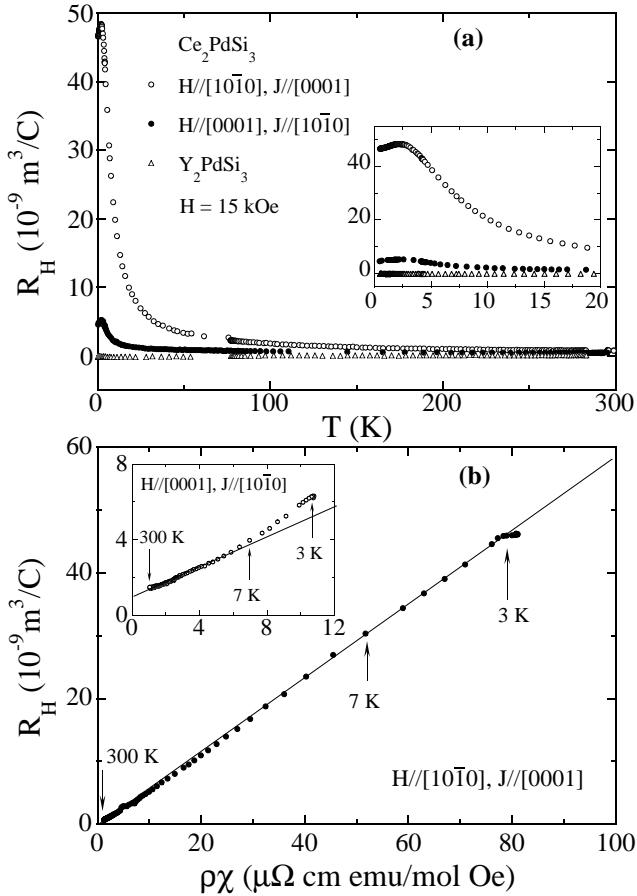}
\caption{(a) The Hall coefficient R$_{H}$ (employing a magnetic field
of 15 kOe) as a function of temperature for Ce$_{2}$PdSi$_{3}$ single
crystals with two different orientations and for polycrystalline
Y$_{2}$PdSi$_{3}$. The inset shows the expanded view of low-temperature
$R_{H}(T)$. (b) R$_{H}$ as a function of the electrical resistivity
times magnetic susceptibility (with temperature as an intrinsic parameter).}
\end{figure}
\end{center}
The $S(T)$ curve for $\Delta T$//[0001] at high
temperatures is almost parallel to that of Y$_{2}$PdSi$_{3}$, therefore
a significant effect of the Pd-4$d$ band on $S(T)$ even in Ce$_{2}$PdSi$_{3}$
cannot be ruled out. The temperature dependence of Hall coefficient ($R_{H}$)
for $H=$ 15 kOe,
shown in Fig. 5(a), also reflects the anisotropic nature of this material. For
both geometries (as labeled in the figure) $R_{H}$ is positive at room
temperature and increases gradually with decreasing temperature. At low
temperatures $R_{H}$ becomes highly anisotropic and shows a positive peak for
both geometries, in the vicinity of $T_{N}$ (see the inset). The large
anisotropy observed in $R_{H}$ is also reflected in the $\chi(T)$ data taken
at $H=$ 15 kOe (not shown), indicating that the anisotropy in $R_{H}$ is of
magnetic origin. The positive value of $R_H$ at all temperatures and a
positive peak at the vicinity of $T_{N}$ are similar to those in the
antiferromagnetic Kondo-lattice compound CeAl$_{2}$.\cite{27} In contrast,
$R_{H}$ ($\simeq$ 0.9$\times$10$^{-10}$ m$^{3}$/C at 300 K) in
Y$_{2}$PdSi$_{3}$ is almost temperature independent. Clearly, there
is a dominant 4$f$ contribution in Ce$_2$PdSi$_3$. The Hall coefficient in
magnetic materials like those in Ce compounds is generally a sum of two
terms; an ordinary Hall coefficient ($R_{0}$) due to Lorentz force and an
anomalous part arising from magnetic scattering (skew scattering);\cite{27}
in the paramagnetic state $R_{H}=R_{0}+A\rho\chi$, where $A$ is a constant.
Using this relation, $R_{0}$ is estimated by plotting
$R_{H}$ versus $\rho\chi$ [Fig. 5(b)]. From Fig. 5(b), it is obvious that the
plot is linear for both $H//[10\overline10]$ and $H//[0001]$ in the
paramagnetic state with a value of $R_0\simeq -$3.2$\times$ 10$^{-10}$
m$^{3}$/C and 1.0$\times$10$^{-9}$ m$^{3}$/C, and $A\simeq$
7.4$\times$10$^{-16}$ mol/C and 4.8$\times$10$^{-16}$ mol/C, respectively.
This linear behavior indicates the presence of dominant skew scattering
in Ce$_{2}$PdSi$_{3}$. In the vicinity of $T_{N}$, however, the data deviate
from the high-temperature linear variation. $R_{0}$ of different sign with the
anisotropic values for two geometries indicates the presence of anisotropy in
the Fermi surface, in agreement with the $S(T)$ data.

\section{Summary}
Summarizing, we have investigated the magnetic behavior
of recently synthesized Ce$_{2}$PdSi$_{3}$ in the single-crystalline
form and the results show strong anisotropic behavior of the
measured properties. The paramagnetic Curie temperature for
$H//[10\overline10]$ is positive, however, the value is negative for
$H//[0001]$. The sign of the thermoelectric power is different for the two
measured crystallographic orientations at high temperatures. Distinct features
due to an interplay between CEF and Kondo effect have also been
observed in the thermoelectric power and resistivity data. The ordinary
Hall-coefficient is anisotropic with opposite sign for the two measured
geometries. The results establish that this compound is an antiferromagnetic
Kondo lattice with $T_{N}$= 3 K. The magnitude of $T_{K}$ is also estimated
to be of same order as $T_{N}$ and this fact suggests a delicate competition
between the Kondo effect and indirect exchange interaction. Therefore it
would be of interest to investigate this compound under high pressure.

\vskip 0.5cm
\begin{center}
{ACKNOWLEDGMENT}
\end{center}

This work has been partially supported by a Grant-in-Aid for Scientific
Research from the Minstry of Education, Science, Sports and Culture of Japan.

\end{document}